# Internet of Things: Current Challenges in the Quality Assurance and Testing Methods

Miroslav Bures[1][0000-0002-2994-7826], Tomas Cerny[2][0000-0002-5882-5502]

and Bestoun S. Ahmed[1][0000-0001-9051-7609]

[1] FEE, CTU in Prague, Karlovo nam. 13, 121 35 Praha 2, Czech Republic
[2] Computer Science, Baylor University, TX, USA
`miroslav.bures@fel.cvut.cz`

**Abstract.** Contemporary development of the Internet of Things (IoT) technology brings a number of challenges in the Quality Assurance area. Current issues related to security, user's privacy, the reliability of the service, interoperability, and integration are discussed. All these create a demand for specific Quality Assurance methodology for the IoT solutions. In the paper, we present the state of the art of this domain and we discuss particular areas of system testing discipline, which is not covered by related work sufficiently so far. This analysis is supported by results of a recent survey we performed among ten IoT solutions providers, covering various areas of IoT applications.

**Keywords:** Internet of Things; Quality Assurance; Testing Methodology; Test Strategy; Integration Testing; Security; Interoperability; Integration Issues

## 1   Introduction

In last two decades, the Internet of Things (IoT) solutions started to emerge from the initial pioneering visions to regular industrial solutions, which are present in our everyday lives. The lively development of these solutions brings also a number of challenges [1,2]; as common examples, we can discuss the insufficient level of standardization, legislation and quality assurance techniques, as well as security and privacy concerns [3,4,5,6]. In this paper, we focus on the quality assurance and testing techniques for the IoT domain. Despite the fact, that some of the areas are intensely covered by the literature (security and privacy are the typical representatives), in the area of systematical testing and quality assurance methodologies, much less work exists. In this paper, we present an overview of the domain and identify the areas, which we consider relevant for the further research. This analysis is supported by discussion of the specifics of IoT solutions having an impact on particular testing techniques and methods, together with a literature survey and with a survey among ten IoT solutions providers, which provided us with different, independent viewpoints on the problem discipline.

    The paper is organized as follows. Section 2 analyzes principal issues of IoT solutions, leading to challenges in IoT quality assurance. Section 3 summarizes the state of the art in this domain. Section 4 presents the results of the recent survey among IoT

solutions providers. In section 5 we discuss the results and we identify the quality assurance areas, which have to be covered by a more intense research. The last section concludes the paper.

## 2   Principal IoT Issues with Impact on Testing Techniques

A number of discussions have been conducted regarding the IoT issues, for instance in [1,2,3,4,5,7,8]; however, during our literature survey, we have not found a systematic analysis, identifying what is the impact of these specifics to particular software testing methods and techniques. Hence, we provide such an analysis in this paper. In the following section, we identify several typical issues of IoT solutions and we number them by IDs. Next, in Table 1, we map these issues with direct consequences they have on the testing and quality assurance process.

**Issue 1.** From the business and economic viewpoint, competition in IoT business is having a direct impact on the conditions, in which these solutions are developed. This competition triggers a demand to lower prices of the manufactured IoT devices, as well as it creates a pressure to shorten time to market.

**Issue 2.** In specific applications of the IoT as the sensor networks or camera networks are, the devices can be located in places, which makes them easily accessible by an attacker; on the other hand, difficult to check by the service provider periodically. These devices can act as a vulnerable point to the entire network.

**Issue 3.** Another related issue is a low possibility to update certain types IoT devices. Either due to low production costs or energy consumption issues it is not possible to update some types of devices, which is typical for sensor networks. This has two consequences: (1) known security defects can be exploited by a potential attacker, and (2) inability to update the device firmware can lead to significant number of various versions of the devices used in production run of the service; these variants need to be tested, which increases the costs of the testbed and also number of variants to test.

**Issue 4.** The IoT devices powered by battery or solar energy lead engineers to minimize the power consumption of the device. This can lead to the implementation of lightweight authorization and security algorithms, exposing these IoT devices as a weak entry point to the whole network.

**Issue 5.** Compared to common web-based internet solutions, testing IoT solutions is specific from another viewpoint. When testing the web-based systems, we usually assume, that the lower physical layers (hardware, network protocols, operational systems, application servers etc.) are tested sufficiently already by supplier parties. Hence, we focus the system testing effort mainly on the application and integration levels. In IoT, the situation is utterly different. Compared to web-based solutions, there is a much more extensive variety of used standardized protocols [9]. Moreover, a number of proprietary protocols are used in the current IoT solutions. Thus, testing IoT services usually involves specific testing of the lower layers of the system; when a service involves development of the own IoT devices, we need to test also this hardware.

**Issue 6.** IoT devices are connected to the Internet network, which has at least two consequences: (1) number of links between connected devices will grow rapidly, and (2) weakly secured device can act as an entry point to the entire network.

**Issue 7.** In a number of IoT devices, the user can have low insight into the internal mechanism of a device; also, if a device is updated, the user can have low control about these updates. Combined with GPS, voice recognition or embedded cameras, this can lead to serious security and privacy threats.

**Issue 8.** Home-made devices not implementing industry standards can be produced and these devices can be integrated together with standardized IoT devices.

**Issue 9.** The dependency of the user to the network service is slowly, but constantly, growing, and this trend has to be expected to continue. In the IoT solutions, this can be especially critical in the case of medical or mission-critical services, where the reliability of the service must be ensured.

More issues can be identified; in this discussion, we tried to identify the most significant potential problems. Table 1 matches the identified issues with their consequences for the system testing processes.

After this initial analysis, let us discuss the IoT quality aspects and techniques, which are currently being researched.

**Table 1.** Consequences of IoT issues for testing methods.

| Issues | Consequence for testing methods |
|---|---|
| 1, 5, 9 | Demand for comprehensive method to define efficient test strategy for IoT solutions |
| 2, 3, 4, 6, 7 | Increased demand for security testing, including privacy aspects |
| 3, 8 | Demand for more efficient methods how to select economic but representative platform variants to test |
| 3, 5, 8 | Increased demand for more efficient integration testing, if possible, automated |
| 1, 3, 5 | Test automation in general, as the number of variants seems not feasible to be tested manually |
| 9 | Testing of behavior of the IoT solutions under limited connection and various edge conditions is needed, especially for life-critical systems |

## 3   Related Work

In the current literature, several principal areas dealing with IoT quality can be identified. We can categorize them as the following: (1) security issues, (2) user's privacy and trust issues, (3) reports on IoT testbeds and (4) other quality assurance and testing techniques not related to security, privacy, and particular testbeds. In this section, we summarize these areas.

In our literature survey, we analyzed selected 371 papers related to the categories above from the IEEExplore, ACM Digital Library, and SpringerLink databases. Papers shorter than 4 pages, technical reports, and popular articles were excluded from the analysis. Table 2 summarizes the numbers of papers related to these categories.

Table 2. Number of papers related to principal categories.

| Category | Number of papers |
| --- | --- |
| Security issues | 261 |
| User's privacy and trust issues | 43 |
| IoT testbeds | 38 |
| Quality assurance and testing techniques | 29 |

In the related literature, **Security issues** are frequently discussed. A number of papers raise the concerns related to security issues, for example [3, 6, 10], and analyze the possible security problems [4, 5]. Moreover, for security testing as a standalone discipline, a number of reports can be found, as an example, we can give [11, 12]. Furthermore, a number of secure architectures on a conceptual and physical level are discussed, for instance [13, 14]. The security area is covered by live publication activity, which reflects on the importance of the issues related to IoT security.

A related topic, user's **privacy and trust** is also being frequently discussed. Concerns are raised [7, 8] and together with that, privacy-aware IoT architectures are being reported [15, 16]. In some of the studies, the privacy and trust topic is overlapping with the security issues, for instance [6, 10].

In the literature, a number of reports on various **IoT testbeds** (or test environments) can be found. Proposed architectures of these testbeds vary from standalone setups [17], distributed architectures [18], or crowd-sourcing based testbeds [19]. Some of the proposals are also based on the simulation of IoT physical devices, e.g. [20], which is a logical step due to the costs of a physical test environment.

The remaining area to discuss is **QA and testing techniques**. This area covers functional testing of IoT solutions, its integration testing, Model-Based Testing and related techniques. Due to the scope of our paper, these reports are the main subject of our interest. Here, we present the more detailed overview.

Several standard-established sub-disciplines of system testing research are spanning to the IoT testing currently. As the initial example, we can give the **Model-Based Testing**. IoT systems are being modeled by a semantic description of IoT services [21] or by several IoT-specific variants of state machines [22]. Also, UML-based models can be found; for instance, UML class and object diagrams are combined with Object Constraint Language [23]. Alternatively, UML Sequence diagrams with Π-calculus are used [24]. From these models, test cases are generated automatically, which increases the accuracy and coverage of these tests. Closely related to the Model-Based Testing, the **Model Checking** discipline has its representatives in the specific IoT context. To detect possible inconsistencies and defects in IoT models, Computation Tree Logic, CTL [25], δ-Calculus [26] or Temporal Logic of Actions (TLA) formal specification language, based on temporal logic [27] are used. The first representatives of the **runtime verification** of the IoT solutions can be found [28]. In this context, we can also

444

The issues were specified as follows. **Limited connection** means behavior of the IoT system under limited network connection. **Interoperability** included mutual compatibility of the IoT devices, missing or insufficient standards and a question of proprietary vs. internet standards. The **number of configurations** means the number of various configurations and types of the end nodes, making the solution hard to test on all these combinations, in software testing, this effect is called "combinatorial explosion."

**Security issues** cover various security breach scenarios, where IoT device serve as a weak entry point to the network, possible security breach leading to a personal harm of the user, or security breach leading to a violation of the user's privacy. Here, the area overlaps with the **Privacy**, which also covers possible misuse of collected personal data and reconstruction of user's digital portrait from various data streams. **Integration** issues include challenges how to test interactions of the individual IoT devices and their behavior in the edge cases, this area also relates to the interoperability of the devices. **Test effort focus** stood for a challenge, how to determine an efficient and specific test strategy for an IoT solution, which would determine the intensity of testing, test levels, and specific testing techniques. **Performance** issue covered behavior of the IoT solution under possible user traffic peeks and various limited conditions (e.g., a combination of the user traffic peek with a limited network connection). Finally, **Legislation** covered various issues related to the necessity to comply with local legislation, or vague definitions of the implementation rules in this legislation.

Regarding the IoT quality issues considered as significant, the results of the survey presented in Table 3 are relatively balanced; rather than pointing out a clear outlier, the data document, that the mentioned aspects are considered important by the industry representatives. Moreover, IoT quality issues considered significant varied by particular business domain of the IoT solution provider.

As the most significant issues, a behavior of IoT solution on a limited connection, interoperability and problems with a number of various versions and platform variants to test have been pointed out, closely followed by security and integration issues.

## 5 Discussion

Considering the related literature covering the principal IoT quality areas (Section 3, Table 2), a discussion can be made, whether integration, interoperability, platform variants and limited connection problems, shall be covered by the more intense development of IoT-specific testing and quality assurance techniques.

A question can be raised, whether the current software and system testing techniques in their generic form are insufficient to ensure proper testing of the IoT solutions. However, from our feedback from the industry survey (Table 3) as well as from our findings in the initial analysis (Table 1), the conclusion suggests, that this area is rather potential for future research.

In this section, let us further discuss three of these areas: (1) interoperability, (2) behavior of IoT solutions on a limited connection and (3) testing problems caused by a number of various versions and platform variants.

The interoperability of various IoT devices can be addressed by IoT-specific testing methods in two lines. The first line raises the current demands on automation of integration testing and simulation of parts of an IoT infrastructure. Consequences go to the Model-Based Testing discipline. Here, path-based or state-machine-based test case generation techniques can be adapted to the IoT-specific context.

The second line focuses on unit-level integration testing and raises demands to select suitable platform variants, also to generate efficient sets of input testing data for this integration tests. This generates an opportunity for the Constrained Interaction Testing discipline.

Also, the behavior of IoT solution under a limited network connection (or other solution-specific limiting constraints) raises the demands for specific integration and end-to-end testing; also, here, Model-Based Testing discipline could provide more specific methods. A possible approach could be modeling the reliability of the particular network lines in the model of the System Under Test and reflection of these specifics in a generation of special test cases addressing this problem.

Finally, a high number of platform configurations and variants to test is the domain of the Combinational Interaction Testing and Constrained Interaction testing disciplines. IoT-specific models for this problem can be created by modification of the current modeling notations (e.g. Combinational Arrays of Feature Models) to allow generation the test cases efficiently addressing the problem.

Also, overlapping with the interoperability issue, increased demand for integration testing of the IoT solutions and automation of these tests opens an opportunity for further development of integration testing frameworks, to decrease potential maintenance of automated tests, frequently reported as the major drawback of this technology [38]. In the area of front-end based automated testing, the maintenance issues are covered by previous work, e.g. [39, 40, 41]. However, this is not the case for the automated integration testing for IoT solutions.

Hence, one of the possible directions here is development of integration testing framework based on unit test framework principles; however, being technically adopted to specifics of the integration test. As an example, we can give higher support for orchestration of an integrated test, more possibilities to chain and execute conditional test steps and more flexible interruption handling of the test flow, all this features also implicitly contributing to decreased maintenance costs of the automated testware.

## 6 Conclusion

Despite the fact, that IoT represents the major and significant stream in the current technology development, related work addressing the topics of IoT-specific testing methods is rather limited. The industry survey presented in this paper documents the demand of the IoT solution providers for efficient testing and quality assurance methods, developed for IoT specific environment.

During our analysis, we have identified three principal areas, which can be the subject of the further research of IoT-specific testing methods: interoperability testing tech-niques, techniques for testing of the behavior of the IoT solution under a limited network connection and techniques to efficiently reduce a high number of platform configurations and variants to test. Also, automated integration testing of IoT solutions is one of the prospective streams to be explored further.

IoT-specific Model-Based Testing is one of the suitable candidates to contribute to this area, moreover, Model Checking discipline can explore possibilities of static testing of IoT designs to minimize design errors in IoT solutions. Due to the present intensive research and development work in the IoT technology, we can expect more methods to be developed by the research community; however, currently, these areas represent further research opportunities.

**Acknowledgements**